\documentclass[12pt]{iopart}
\usepackage{graphicx}
\usepackage{bm}
\begin{document}
\title{Anomalous scaling and universality in hydrodynamic systems with
power-law forcing}

\author{L.~Biferale\dag,
M.~Cencini\P\footnote[2]{To
whom correspondence should be addressed (massimo.cencini@roma1.infn.it)}, 
A.~S.~Lanotte\S, M.~Sbragaglia\dag and F.~Toschi$\|$}

\address{\dag\ Dip. di Fisica and INFM, Universit\`a "Tor Vergata", Via
della Ricerca Scientifica 1, I-00133 Roma, Italy}
\address{\P\ Center for Statistical Mechanics and Complexity,
INFM Roma ``La Sapienza'' and Dipartimento di Fisica, Universit\`a di
Roma "La Sapienza", Piazzale Aldo Moro, 2, I-00185 Roma, Italy}
\address{\S\ Istituto per le Scienze dell'Atmosfera e del Clima,
CNR, Str.~Prov.~Lecce-Monteroni km.~1.200, I-73100 Lecce, Italy; and INFM UDR Tor Vergata}
\address{$\|$ Istituto per le Applicazioni del Calcolo, CNR, Viale
del Policlinico 137, I-00161 Roma, Italy, and INFM UDR Tor Vergata}
\begin{abstract}
The problem of the interplay between normal and anomalous scaling in
turbulent systems stirred by a random forcing with a power law
spectrum is addressed. We consider both linear and nonlinear
systems. As for the linear case, we study passive scalars advected by
a $2d$ velocity field in the inverse cascade regime. For the
nonlinear case, we review a recent investigation of $3d$
Navier-Stokes turbulence, and we present new quantitative results for
shell models of turbulence.
We show that to get firm statements is necessary to reach
considerably high resolutions due to the presence of unavoidable
subleading terms affecting all correlation functions.  All findings
support universality of anomalous scaling for the small scale
fluctuations.
\end{abstract}
\bigskip
\maketitle
\section{Introduction}
\label{sec:intro}
The understanding of the small scale statistics of turbulent systems
is a problem of considerable interest \cite{frisch}. With turbulent
systems, we mean both the dynamics of velocity fields in high-Reynolds
number flows, and the advection of scalar or vector fields by
turbulent flows as, e.g., temperature or magnetic fields. In the last
few years, much progress has been reached both in experimental
\cite{NWLMF97,LPVCAB01,MMMP01,W00} and numerical \cite{GFN02,BCV00}
investigations of turbulent systems. We have now plenty of
observations showing that two properties generally holds.  First,
turbulent systems are intermittent \cite{frisch,SA97,W00}, as
quantified by the anomalous scaling of moments of field increments.
Second, the anomalous scaling exponents display universality with
respect to the boundary conditions and to the {\it large scale}
forcing mechanisms \cite{SA97,ABBBC96}. Some subtle points arise when
isotropy is broken at large scale by the injection mechanisms
\cite{ADKLPS98,W00,KS00,SW02}. In that case, for universality to hold
it is required that anisotropic contributions are subleading with
respect to the isotropic one. However, evidences for universality of
both isotropic and anisotropic scaling exponents have been found
\cite{BT00,BCTT03}.

Nonetheless, the mechanisms responsible for anomalous scaling and
universality in turbulent systems are still not fully comprehended with
the remarkable exception of linear problems such as the passive
transport of a field, for which intermittency and universality have
been systematically understood (see Ref.~\cite{FGV01} for an
exhaustive review). In this context, in particular for the class of
Kraichnan models \cite{K68}, closed equations for the correlation
functions can be derived. These are linear partial differential
equations, whose homogeneous solutions (zero modes) generally exhibit
anomalous scaling. On the other hand, the inhomogeneous solutions,
constrained by the external forcing, possess dimensional
(non-anomalous) scaling. Universality results then from the decoupling
between the zero modes scaling and the forcing. Remarkably, the zero
modes can be interpreted as statistically preserved structures,
i.e. functions that do not change in time once averaged over the
velocity field realisations and particle trajectories. This allowed
for successfully testing the entire picture also for the advection by
realistic velocity fields \cite{CV01,CLMV01}, and in the context of
shell models for passive transport \cite{ABCPV01}.

On the theoretical side, it is tempting to export the concepts issuing
from linear turbulent systems to nonlinear ones, i.e.  to see whether
the same mechanisms for anomalous scaling and universality are at work
also in nonlinear hydrodynamic systems such as Navier-Stokes
turbulence.  A possible test case to probe universality with respect
to the forcing mechanisms is to study turbulent fluctuations stirred
at all scales by a self-similar power-law random forcing. Indeed, the
non-analytical properties of the forcing may alter the energy exchange
between the turbulent field fluctuations. For instance, in some cases,
the energy injection mechanism may prevail over the energy cascade
process, modifying the inertial range physics.

This problem was pioneered by means of Renormalisation Group (RG)
methods \cite{FNS77,FF78,DDM79}, in the late '70s, for the
$d$-dimensional Navier-Stokes equations. In the RG calculations, for
$d=3$, the forcing spectrum is chosen as $E_{f}(k) \sim k^{3-y}$ with
$y$ playing the role of a small parameter in perturbative
expansions. Unfortunately the interesting physical case,  obtained
for $y=4$ and corresponding to the Kolmogorov spectrum for the velocity
field, lies in a range where convergence of the RG expansion is not
granted \cite{EY94}. Notwithstanding extensions of the RG formalism to
$y \sim O(1)$ values have been tempted by different approaches
\cite{YO86,AAKV03}, the problem is still open. Recent numerical
simulations tried to shed light on this issue \cite{SMP98,BLT03} but,
because of the limited resolution, their results are not conclusive.

The aim of this paper is to investigate the general issue of the small
scale statistical properties of linear and nonlinear hydro-dynamical
systems, in the presence of stirring acting at all scales with a
power-law spectrum. A systematic study of the scaling behaviour at
varying the forcing spectrum allows indeed for understanding the
interplay of dimensional and anomalous scaling in turbulent fields.

In Sec.~\ref{sec:ps}, as an instance of linear problems, we consider
passive scalars stirred at all scales by a power-law forcing, and
advected by a $2d$ turbulent velocity field.  Our results confirm the
universality scenario originating from the zero modes picture which
predicts two distinct regimes. {\it Forcing-dominated} regime: the
scaling of low order structure functions is non-anomalous, with
exponents dimensionally related to the forcing spectrum; for the
higher order moments, scaling is anomalous and dominated by the zero
modes. {\it Forcing-subleading} regime: the dimensional scaling
related to the balance with the forcing is subleading, at any order,
with respect to the anomalous one, similarly to the case of a standard
large scale injection.  Hence, anomalous scaling is observed for any
order statistics.  A subtle technical point revealed by the passive
scalar case is the existence of many important power-law terms that
contribute to the scaling properties. As clarified in
Sec.~\ref{sec:ps}, for both regimes, to disentangle the authentic
scaling behaviours, it is necessary to take into account the leading
as well as the subleading terms.  Concerning nonlinear systems, in
Sec.~\ref{sec:NL}, we review the numerical study of the $3d$
Navier-Stokes equations done in Ref.~\cite{BLT03}, where due to
natural limitation in the resolution only semi-quantitative results
were obtained. Then, to overcome this difficulty, we consider shell
models for turbulence.  These are nonlinear models that maintain most
of the richness of the original problem but allow for reaching higher
Reynolds numbers. Shell model results coherently fit with the passive
scalar and NS ones.  Conclusions follow in Sec.~\ref{sec:conclude}.
\section{Linear dynamics: passive scalar transport}
\label{sec:ps}
The evolution of a passive scalar field $\theta$ advected by an
incompressible flow ${\bm v}$ is governed by the advection-diffusion
equation\,:
\begin{equation}
\partial_t \theta \,+\, {\bm v} \cdot {\bm \nabla}\,\theta\,=\,\kappa\,\Delta\,
\theta \,+\,f\,,
\label{eq:ps}
\end{equation}
where $\kappa$ is the molecular diffusivity.  The standard
phenomenology is as follows. Scalar fluctuations are injected at
large scale by the source term $f$; then, by a cascade mechanism
induced by the advection term, they reach small scales where
dissipation takes place balancing the input, and holding the system in
a statistically stationary state.

Since the problem is passive, it can be studied also with a synthetic
velocity field mimicking some of the features of realistic turbulent
flows. Much attention has been recently devoted to the Kraichnan model
\cite{K68}, where ${\bm v}$ is a Gaussian, incompressible,
homogeneous, isotropic and $\delta$-correlated in time field of zero
mean. The only reminiscence of real turbulent flows is the existence
of a scaling behaviour that, due to the assumed self-similarity, is
fixed by a unique exponent, i.e. $\langle (\delta_r {\bm v} \cdot
\hat{\bm r})^2\rangle \sim r^{\xi}$ with $0<\xi <2$ (where $\delta_r
{\bm v}= {\bm v}({\bm x}+{\bm r})-{\bm v}({\bm x})$). For the sake of
analytical control, the forcing $f$ is also taken as a Gaussian,
homogeneous, isotropic field of zero mean and with correlation function:
$\langle f({\bm r},t) f({\bm 0},0) \rangle=\delta(t) {\cal F}(r)$,
where ${\cal F}(r)$ decays rapidly for $r\gg L_f$, identifying $L_f$
as the large scale of the problem.

Thanks to the above simplifying assumptions, a closed equation for the
generic $p$-point correlator, $C_{p}({\bm r},t)=\langle \theta({\bm
  r}_1,t) \theta({\bm r}_2,t)...\theta({\bm r}_p,t)\rangle$, can be
derived. It formally reads
\begin{equation}
 \partial_t C_{p} = - {\cal M}_p C_{p} + {\cal F} \otimes C_{p-2}\,,
\label{eq:correlaps}
\end{equation}
where ${\cal F}$ is the forcing spatial correlation defined above, and
${\cal M}_p$ is a linear differential operator coming from the
diffusion and the advection terms (see Ref.~\cite{FGV01} for details).
The stationary solution of (\ref{eq:correlaps}) satisfies the equation
${\cal M}_p C_{p} = {\cal F} \otimes C_{p-2}$.  As usual, a linear
equation is solved by the superposition of the solution of the
associated homogeneous equation, ${\cal M}_p {\cal Z}_{p}=0$, and the
solution of the inhomogeneous one, which will be denoted as
$C_{p}^{\cal I}$. In the inertial range of scales both $C_{p}^{\cal
I}$ and ${\cal Z}_{p}$ display a scaling behaviour, i.e.
\begin{equation}
\fl C_{p}^{\cal I}(\lambda \underline{\bm r})\sim
\lambda^{\zeta^{dim}_p}C_{p}^{\cal I}( \underline{\bm r})
\qquad {\mbox {and}} \qquad
 {\cal Z}_{p}(\lambda \underline{\bm r})\sim
\lambda^{\zeta_p} {\cal Z}_{p}( \underline{\bm r})\,,
\qquad {\mbox {with}}\; \underline{\bm r}=({\bm r}_1,\dots,{\bm r}_p)\,,
\end{equation}
where the scaling exponents $\zeta_p^{dim}=p(2-\xi)/2$ are fixed by
dimensionally matching the inertial operator with the forcing, while
the $\zeta_p$, not constrained by any dimensional requirements, are
independent of the forcing. One is usually interested in the
$p$-point irreducible component of the correlation function, $C_p$,
that is the structure function $S_p(r)=\langle (\delta_r
\theta)^p\rangle$. Still in the framework of Kraichnan model, the zero
modes scaling exponents contributing to $S_p$ have been calculated
perturbatively in \cite{GK95,CFKL95}. It has been found that
$\zeta_p<\zeta_p^{dim}$ for $p>2$, i.e. the leading contribution to
the structure function scaling is anomalous.

The zero modes dominance scenario explains both the anomalous scaling
exponents and their universality. Indeed the forcing does not enter
the definition of ${\cal Z}_p$, but fixes only the multiplicative
constants in front of $C^{\cal I}_{p}$ and ${\cal Z}_p$, necessary to
match the large scale boundary conditions. This implies that, even
though scaling exponents are universal, probability density functions
(PDF) of scalar increments are not.  The above picture based on zero
modes is now recognised to apply in all linear hydrodynamic
problems. As confirmed by investigations of passive advection by
realistic velocity fields \cite{CLMV01,CV01}, and of shell models for
passive transport \cite{ABCPV01}.

Let us now focus on the main issue of this work: the scaling behaviour
in the presence of forcing fluctuations directly injected in the
inertial range of scales. We consider the case of a
source term $f({\bm r},t)$ which is a random Gaussian scalar field,
with zero mean, white-in-time and characterised by the spectrum
\begin{equation}
E_f(k)=\pi k \langle |f({\bm k})|^2\rangle \propto k^{-1+\beta}\,. 
\label{def:force_ps}
\end{equation}
In the following, it is always assumed the presence of ultraviolet and
infrared cutoffs for the forcing, i.e. expression (\ref{def:force_ps})
holds only in the range $k_1\,<\,|{\bm k}|\,<\,k_2$ of wave-numbers,
where $k_1$ and $k_2$ are of the order of the inverse of the largest
scale in the system and of the dissipative scale, respectively.

 Remaining in the framework of the Kraichnan model,
Eq.~(\ref{eq:correlaps}) still holds, and its stationary solution is
again the superposition of the homogeneous solution, ${\cal Z}_p$,
which remains unchanged ({\it viz.} with the same scaling exponents)
and of the inhomogeneous one, $C^{\cal I}_p$, whose scaling exponents
now depend on the slope of the forcing spectrum, $\beta$.  Indeed by
dimensional reasoning, we have $\zeta_p^{dim}(\beta)=
p(2-\xi-\beta)/2$.

Two regimes can be identified. If $\beta<0$, the scaling of the
inhomogeneous solution is always subleading with respect to the
anomalous one. The scaling exponents measured from the structure
functions are the same as those obtained with the standard large scale
forcing.  On the other hand, for $\beta>0$, there exists a
critical order $p_c$ such that for $p<p_c$: $\zeta_p^{dim}(\beta) \leq
\zeta_p$, and for $p>p_c$: $\zeta_p \leq \zeta_p^{dim}(\beta)$. In
other words, the scaling behaviour of the low-order structure
functions is non-anomalous and dominated by the forcing. The appearing
of anomalous scaling for $p>p_c$ can be understood due the fact that
$\zeta_p$ as a function of $p$ is concave. Moreover, it is known
numerically \cite{CLMV01} and to some degree analytically \cite{BL98}
that in the Kraichnan model the scaling exponents above a certain
order saturate to a finite value, $\zeta_{\infty}$, which makes even
more intuitive the existence of a finite $p_c$. Some subtle points may
arise for positive $\beta$ larger than $1$. In such a case, the strong
UV components of the forcing spectrum may allow a matching with zero
modes, disregarded for a large scale forcing \cite{BGK98}, exploding
at small scales. Finally, the case $\beta=0$ is marginal, because the
exponents $\zeta_p^{dim}$ coincide with the large scale prediction, up
to possible logarithmic corrections.

Checking the validity of the above predictions in numerical
simulations of a realistic flow is interesting for two reasons. First,
it is a further demonstration, in the Eulerian framework, of the zero
modes picture for anomalous scaling and universality in linear
problems, which was previously assayed with Lagrangian studies
\cite{CV01}. Second, it offers a controlled testing ground for
interpreting some aspects of the nonlinear hydrodynamics which will be
considered in the next section.

As an example of realistic velocity field, we consider two-dimensional
incompressible Navier-Stokes equations in the inverse cascade regime
\cite{PT98,BCV00,T02}:
\begin{equation}
\partial_t {\bm v} + {\bm v} \cdot {\bm \nabla}\,{\bm v} = -{\bm
  \nabla} P + \eta \Delta {\bm v} -\alpha {\bm v} + {\bm f}_v\,.
\label{eq:2dns}
\end{equation} 
The terms in the r.h.s. of Eq.~(\ref{eq:2dns}) have the following
meaning: $P$ is the pressure field; ${\bm f}_v$ injects kinetic energy
at scale $L_0$; $\nu \Delta {\bm v}$ (where $\nu$ is the viscosity)
dissipates enstrophy at small scale; $-\alpha {\bm v}$ removes energy
at the large scales allowing for a statistically stationary state. In
the inverse energy cascade regime, the velocity statistics is
self-similar (not-intermittent), with $\langle(\delta_r{\bm
v}\cdot\hat{\bm r} )^p\rangle\sim r^{p/3}$, but temporal correlations
are non trivial. Moreover, precise numerical measurements of passive scalars
with large scale forcing \cite{CLMV01,CV01} have shown that:  $\zeta_2=2/3$
while  for $p>2$ the exponents are anomalous, and saturation
is observed for $p \geq 10$ with $\zeta_{\infty}\simeq 1.4$.  It is
noteworthy that saturation seems to be generic in passive scalars, as
found also in experiments \cite{MWAT01}.  From a physical point of
view, it means that the strongest scalar fluctuations are
statistically dominated by front-like structures, i.e. huge
variations of the field $\theta$ at very small scales.

Once a power law forcing (\ref{def:force_ps}) is considered, the
predicted dimensional scaling is $\zeta_p^{dim}(\beta)=
p(2/3-\beta)/2$. Therefore, the two above described regimes 
--statistics dominated by zero modes or by the forcing-- should appear
for $\beta<0$ and $\beta>0$, respectively.
\begin{figure} [t!]
\begin{center}
\includegraphics[draft=false, scale=0.4,clip=true]{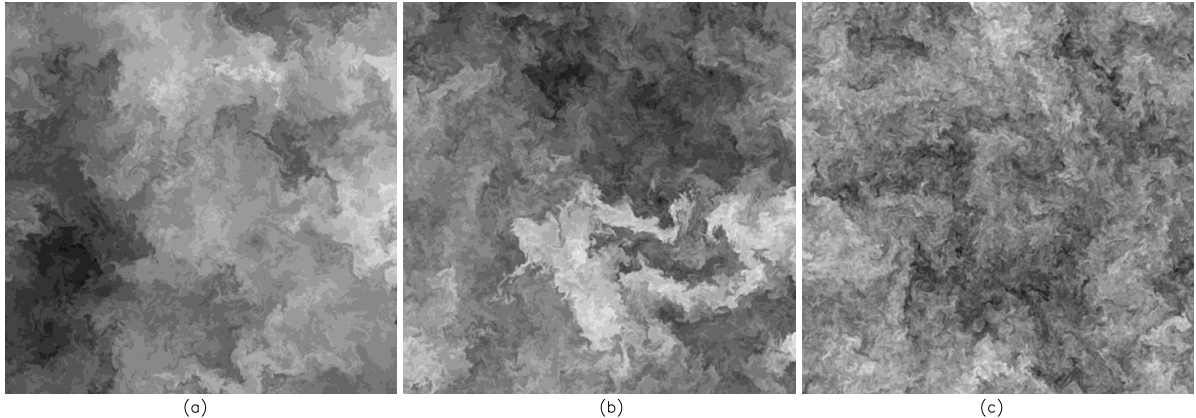}
\end{center}
\caption{Instantaneous snapshots of the scalar field $\theta$ for runs
(a), (b), and (c); intensities are coded in grey-scale. Note that (a)
and (b) are qualitatively very similar, while (c) is characterised by
richer fluctuations at the small scales. The three snapshots are
obtained by integrating Eqs.~(\ref{eq:ps}) and (\ref{eq:2dns}) by
means of a standard $2/3$-dealiased pseudospectral code in a doubly
periodic square domain $2\pi \times 2\pi$ with $1024^2$ grid
points. As for the velocity field, in Eq.~(\ref{eq:2dns}) the viscous
term has been replaced by a hyper-viscous term of order $8$, in order to 
force at the very small scales. 
The velocity forcing ${\bm f}_v$ is chosen as a Gaussian,
incompressible, homogeneous, isotropic and $\delta$-correlated in time
$2d$ field, of zero mean and concentrated around the small scale $L_0$
(of the order of few grid points). The friction coefficient $\alpha$
is tuned in such a way that energy is removed at a scale
$\eta_{fr}\sim \epsilon_{v}^{1/2} \alpha^{-3/2}$ of the order of the
box size. In Eq.~(\ref{eq:ps}),  to have a large inertial
range, the diffusive term has been replaced by a bilaplacian with
$\kappa$ tuned to have the dissipative scale $r_d \geq L_0$. As for
the scalar, in run (a) we used a white-in-time, random Gaussian
forcing concentrated at scale $L_f$ of the order of the box size. In
the runs (b) and (c) we used the forcing (\ref{def:force_ps}) by
choosing $k_1=2$ and $k_2$ in such a way that $k_2 \leq 2\pi/r_d$. For
the sake of comparison, we imposed in all runs the same scalar energy
input. For each run, we collected $80$ frames separated by about one
large scale eddy turnover time measured as $\eta_{fr}/\sqrt{\langle
v^2 \rangle}$. See \protect{\cite{BCV00,CLMV01}} for more details on
the numerical procedure.}
\label{fig:1} 
\end{figure}
We performed three sets of direct numerical simulations (DNS) of
Eqs.~(\ref{eq:ps}),(\ref{eq:2dns}): for run (a), we used a standard
large scale forcing; for runs (b) and (c), we considered a forcing as
in Eq.~(\ref{def:force_ps}) with $\beta=-0.3$ and $\beta=0.3$,
respectively. More details on the DNS are given in the caption of
Fig.~\ref{fig:1}, where we show the snapshots of the scalar field
$\theta$ for all runs. Already at a first sight, it is possible to
observe that runs (a) and (b), where the zero modes dominate over the
inhomogeneous solutions at any order, display qualitatively similar
features: the field is organised into large scale structures or {\it
plateaux} characterised by a good mixing, separated by sharp fronts.
Differently in run (c), which corresponds to the case of forcing
dominated statistics (with $p_c \geq 6$), the scalar fluctuations
develop a wider range of scales, and the most evident large scales
plateaux disappear.
\begin{figure} [t!]
\begin{center}
\includegraphics[draft=false, scale=0.7,clip=true]{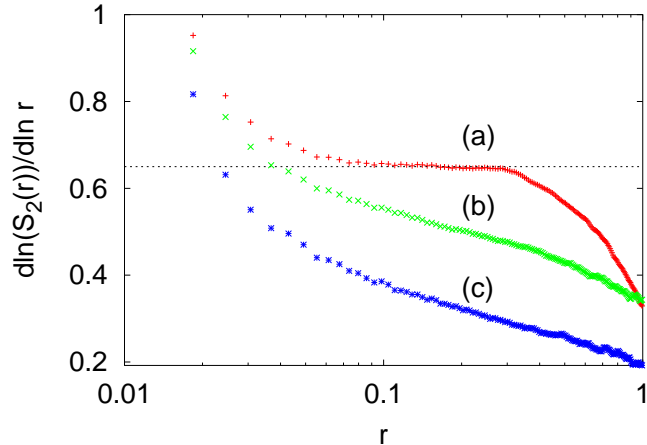}
\end{center}
\caption{Log-lin plot of local slopes of the second order structure
function for the three runs. The dotted straight line indicates the
value $2/3$. Note that in run (a) the slope are constant for about one
decade. On the other hand, the slopes for run (b) and (c) are non
constant in the entire range.  It is worth remarking that for
power-law forcing there are two sources of finite size effects. The
first is the presence of two power-laws, see text. The second is that
the scaling in real space of the forcing two-point correlation is
strongly affected by corrections (induced by the Bessel function) due
to fact that we generate the forcing in Fourier space.  }
\label{fig:2} 
\end{figure}

In order to make the above observations more quantitative, we studied
the scaling of the structure functions $S_{2p}(r)=\langle(\delta_r
\theta)^{2p} \rangle$ (odd orders are zero due to the isotropy of the
forcing). For run (a), we found a very good scaling range of about one
decade, which allowed us for accurately measuring $\zeta_p$ up to
$p\approx 10-12$. The quality of the local slopes and the measured
values agree with previous investigations \cite{CLMV01}, which were
performed with larger statistics and higher resolution.  For the
power-law cases, runs (b) and (c), we could not find evidence of a
good scaling as in run (a). This is observed already from the second order
structure function $S_2$, whose local slope is plotted in
Fig.~\ref{fig:2} for the three runs.

We understand the poor scaling behaviour as due to the competition
between the scaling of the anomalous part and the forcing dominated
one. This becomes clear by looking at the scalar flux,
$\Pi_{\theta}(r) \equiv \langle [\delta_r {\bm v} \cdot \hat{\bm r}]
(\delta_r \theta)^2 \rangle$.  For this quantity under rather general
hypothesis, an exact analytical prediction can be derived for a
generic forcing, through the Yaglom equation \cite{MY75}. For a
power law forcing $f$ as in (\ref{def:force_ps}), we obtain:
\begin{equation}
\label{def:yaglom}
\Pi_{\theta}(r)\equiv \langle [\delta_r {\bm v} \cdot \hat{\bm r}]
(\delta_r \theta)^2 \rangle \sim c_1 r + c_2
r^{1-\beta}\,,
\end{equation}
where we have kept only the first two leading contributions. It is important
to notice that  
constants $c_1$ and $c_{2}$, which depend on the space dimension and
on the details of the forcing spectrum, turn out to have opposite signs.  In
(\ref{def:yaglom}), in addition to the standard linear term, there is
the first leading term induced by the forcing (\ref{def:force_ps}). 
As one can see, the critical value $\beta=0$ naturally arises in the
flux expression to separate the inertial dominated ($\beta<0$) from
the forcing dominated ($\beta>0$) regime. Indeed, for $\beta<0$ the
spectral flux is dominated by the low wavenumber components of the
forcing spectrum and saturates, in Fourier space, to a constant value
as a function of $k$. Scalar fluctuations are transferred down-scale
{\it via } an intermittent cascade and the statistics is dominated by
the anomalous scaling, the forcing contribution even if present at all
scales is subleading. For $\beta>0$, the scalar spectral flux no
longer saturates to a constant: the direct input of energy from the
forcing mechanism affects inertial range statistics in a self-similar
way, down to the smallest scales where dissipative terms start to be
important.

\begin{figure} [htb]
\begin{center}
\includegraphics[draft=false, scale=0.43,clip=true]{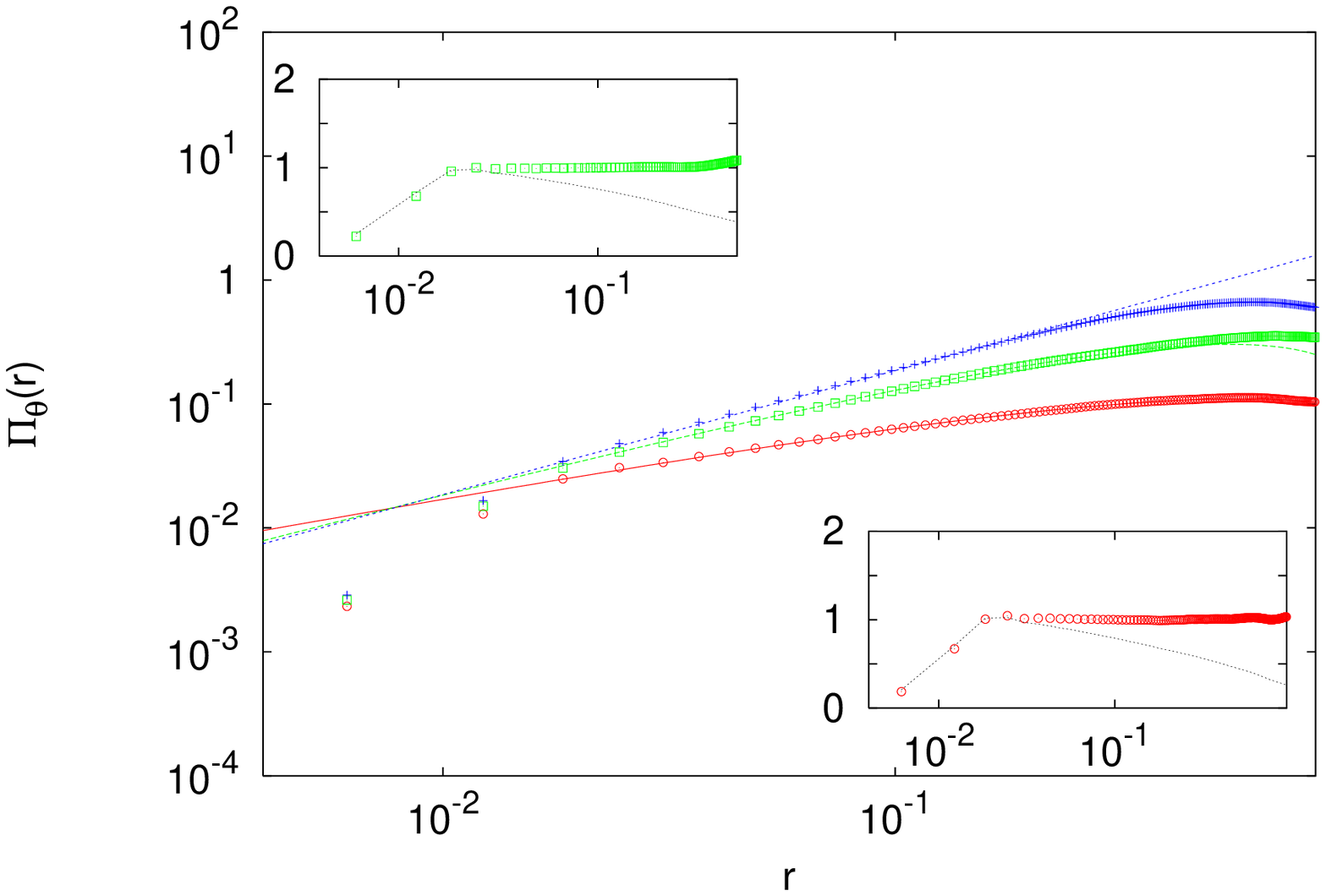} \includegraphics[draft=false, scale=0.43,clip=true]{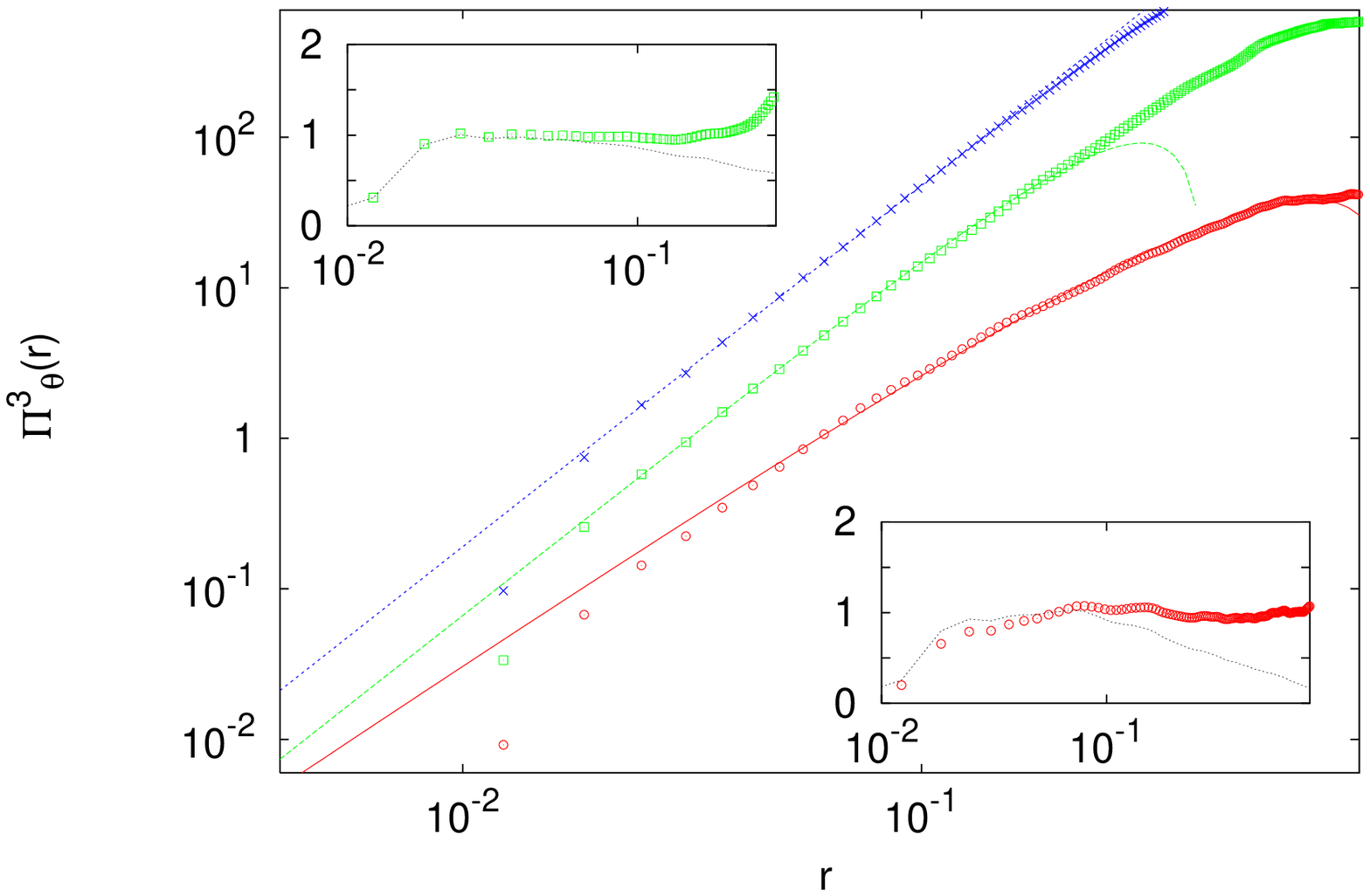}
\end{center}
\caption{Left: structure functions of passive scalar flux variables.
Blue crosses: run (a); green squares: run (b); red circles: run (c).
Fits are made in the inertial range with the expression given by the
Yaglom equation (\ref{def:yaglom}). Insets: same analysis is performed
for compensated ratios in the case of run (b) (top inset) and run (c)
(bottom inset).  We show the best compensation obtained with both
power laws (same symbols as in the body of the figure) in
Eq.~(\ref{def:yaglom}), and with only the leading term (dotted lines).
Right: the same of left panel for the third order moment of the flux
variable. For the large scale forcing we use the single power-law,
$r^{2.4}$, to fit the data, while for the other two cases we used a
power law superposition of the two terms of
Eq.~(\ref{def:yaglom3}). For all fits with a power-law superposition,
the two terms turns out to have opposite signs. Insets: compensations
with a single (leading) or two power laws for run (b) (top inset) and
run (c) (bottom inset).  Here also the best compensation is obtained
with two power laws.}
\label{fig:3} 
\end{figure}

In addition to $\Pi_{\theta}(r)$, we also studied the third moment of
the flux variable $\Pi^3_{\theta}(r)= \langle [(\delta_r {\bm v} \cdot
\hat{\bm r}) (\delta_r \theta)^2]^3\rangle$ whose scaling is anomalous
for the large-scale forcing case,
 i.e. $\Pi^{3}_{\theta}(r)\sim r^{2.4}$ (see also
\cite{CLMV01}). Results are summarised in Fig.~\ref{fig:3}.

Concerning the flux $\Pi_{\theta}(r)$, for the large scale case run
(a), the scaling is good and the Yaglom relation,
$\Pi_{\theta}(r)\approx -2\epsilon_{\theta} r$ (being
$\epsilon_{\theta}$ the scalar dissipation), holds for about one
decade (Fig.~\ref{fig:3}, left panel). On the other hand, for the
power-law forced cases, the scaling is poorer, making the
identification of the exponents less trivial. Indeed, a scaling
behaviour can be properly identified only by taking into account both
the leading and subleading terms in Eq.~(\ref{def:yaglom}). The third
order moment is affected by the same problem. So that, on the basis of
relation obtained for $\Pi_{\theta}$, we tried to fit it with the
following scaling ansatz:
\begin{equation}
\label{def:yaglom3}
\Pi^3_{\theta}(r) \sim A r^{2.4} +B r^{3(1-\beta)},
\end{equation}
where the exponent $2.4$ in the first term is associated to the
anomalous contribution as measured from run (a). Notice that for run
(b) the exponent induced by the forcing is subleading,
i.e. $3(1-\beta)=3.9$, while for run (c) it becomes leading
$3(1-\beta)=2.1$. From the insets of Fig.~\ref{fig:3}, it is clear
that a well defined scaling behaviour is recovered only after having
taken into account the two power-law contributions.  In all cases the
anomalous scaling exponents have values in agreement with those
obtained for a large scale forcing, although the interplay of
different power laws, when the the forcing is as (\ref{def:force_ps}),
makes the identification of the exponents very difficult even at
considerably high resolution.
\begin{figure} [thb]
\begin{center}
\includegraphics[draft=false, scale=.42, clip=true]{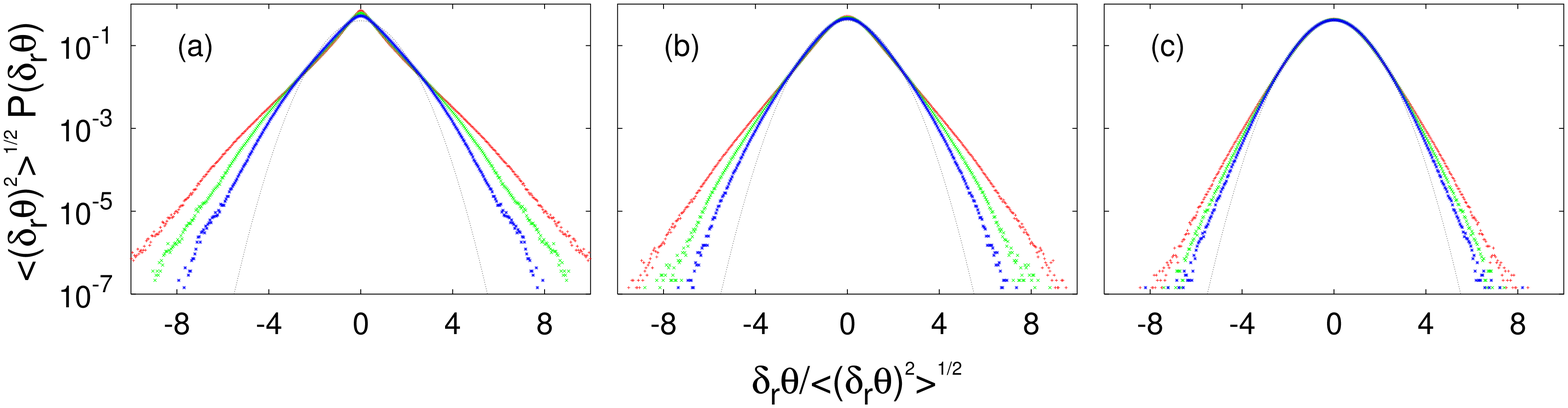}\\
\includegraphics[draft=false, scale=.42,clip=true]{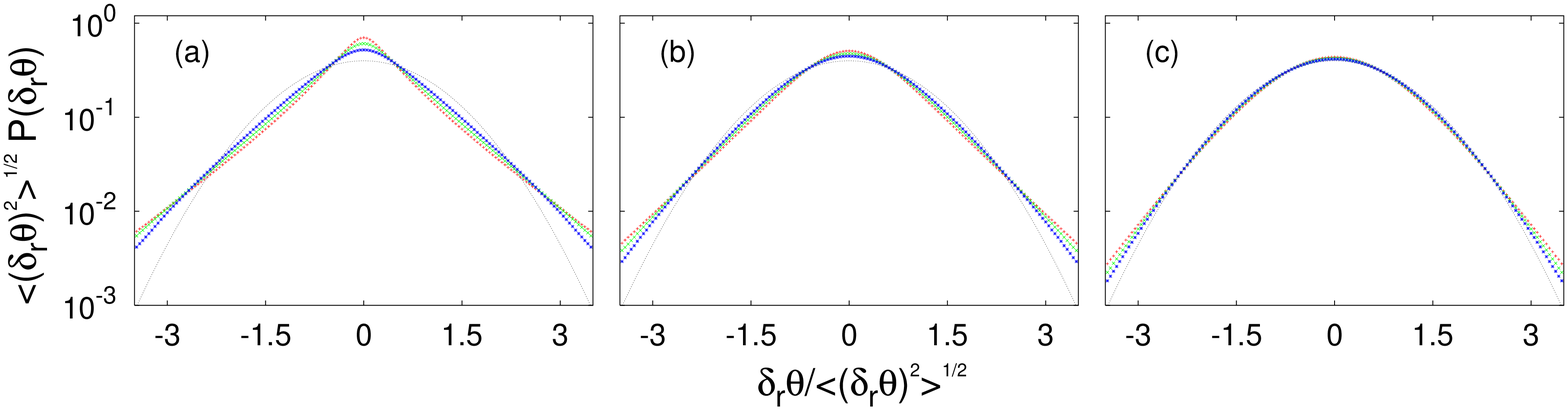}
\end{center}
\caption{Top: PDF of scalar increments, $\delta_r \theta$, for runs
(a), (b) and (c). For each run, we plot the PDF for three different
separations $r$ inside the inertial range. The curves have been
normalised to have unit variance. The dotted line is a Gaussian
distribution with unit variance for comparison. Bottom: the same but
zooming in, to highlight the core behaviour. Note that for run (c) the
PDF's core approaches a Gaussian with a weaker, if not absent,
dependence of $r$.}
\label{fig:4} 
\end{figure}

Due to these difficulties in measuring the exponents, we looked
directly at the PDF of scalar increments $P(\delta_r \theta)$. In
Fig.~\ref{fig:4}, we show plots of normalised PDFs of $\delta_r
\theta$ for different choices of the scale $r$ in the inertial
range. For runs (a) and (b), the PDFs do not display any rescaling
property at changing the scale $r$, neither in the core nor in the
tails, suggesting intermittent behaviour in the full statistics. Note
also that the PDFs in the two runs are different (only scaling
exponents are universal while PDFs are not). Conversely, for run (c),
where the forcing is dominant, the PDFs display a fairly good
rescaling in the core (which governs the low order statistics), while
the tails still do not collapse. This agrees with the existence of a
critical order, $p_c$, above which anomalous scaling appears even in
the forcing-dominated case.
\begin{figure} [htb]
\begin{center}
\includegraphics[draft=false, scale=.7, clip=true]{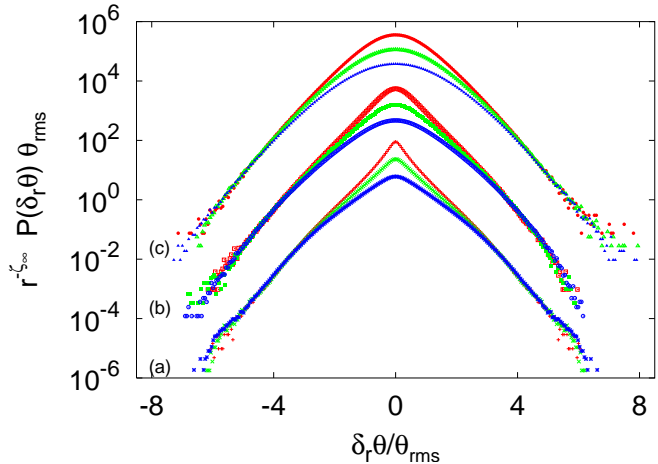}
\end{center}
\caption{PDF of scalar increments rescaled to test
Eq.~(\protect{\ref{def:satura}}) with $\zeta_{\infty}\simeq 1.4$ for all
runs (a), (b) and (c). For plotting purposes, curves of runs (b)  and
(c) have been shifted by a factor $10^3$ and $10^6$, respectively.}
\label{fig:5} 
\end{figure}
As discussed earlier, all differences associated with the two regimes
should disappear in the high order statistics, i.e.  for $p>p_c$ when
anomalous scaling exponents imposed by the zero modes should show up
irrespectively of the forcing. In testing this point, the presence of
saturation for large $p$ comes into help, indeed it entails that, for
large excursions, the PDFs must approach the form \cite{CLMV01}:
\begin{equation}
\label{def:satura}
P(\delta_r \theta)=r^{\zeta_{\infty}} Q\left(\frac{\delta_r
\theta}{\theta_{rms}}\right)\frac{1}{\theta_{rms}},
\end{equation}
with $\theta_{rms}=\sqrt{\langle \theta^2 \rangle}$ and
$\zeta_{\infty}\simeq 1.4$.
Such a prediction is tested and confirmed by the collapse in the PDF
tails shown in Fig.~\ref{fig:5}, which tells us that $\zeta_{\infty}$
is asymptotically approached independently of the forcing.

The above results provide support to the validity
of the zero modes picture beyond the boundaries of the Kraichnan model
and in agreement with the Lagrangian investigations \cite{CV01}.

\section{Nonlinear systems}
\label{sec:NL}
Observations from the linear problem strongly indicate that the
notions of anomalous scaling and universality are closely linked. When
we enter  the nonlinear world, we do not have any longer  a
reference theory. Here, we shall rather formulate some
working hypothesis suggested by the results from the linear systems
and verify their consistency with two cases studied, namely: $3d$
Navier-Stokes turbulence, reviewing the work first presented in
\cite{BLT03}, and shell models \cite{BJPV98} for turbulence.
\subsection{The $3d$ Navier-Stokes problem}
\label{sec:NS}
In the $3d$ case, we consider random forcing defined by a two-point
correlation function that in Fourier space reads
\begin{equation}
\label{def:force}
\langle f_i({\bm k},t) f_j({\bm k}',t') \rangle \propto
k^{1-y}P_{ij}({\bm k}) \delta({\bm k}+{\bm k}')\, \delta(t-t')\,,
\end{equation} 
where $P_{ij}({\bm k})$ is the projector assuring incompressibility
and the forcing spectrum behaves as $E_f(k) \sim k^{3-y}$.
Hereafter, we use the definition $y=4-\beta$, referring to the
classical notation of the problem as introduced in \cite{FNS77,FF78}.
Correspondingly, the critical value $\beta=0$, separating the two
regimes in the linear case, is now $y_c=4$.

The relative importance of the stirring on the small scales can be
varied by tuning the slope value $y$ from $y \sim 0$ (meaning strong
input at all scales) to $y \rightarrow \infty$ (corresponding to a
quasi large scale forcing).  Also in NS turbulence, a physical insight
can be taken from the behaviour of the energy flux, which is constant
through scales up to logarithmic corrections for the subleading
forcing case, $y>y_c$; while it becomes a scale dependent function for
$y<y_c$, overcoming the energy cascade mechanism.

The case of strong input at all scales, $y \sim 0$, was originally
investigated in Ref.~\cite{FNS77} by means of a RG approach, leading to the
following expression for the energy spectrum $E(k)\sim k^{1-2y/3}$, in
the domain $\eta \ll k^{-1} \ll L_f$ where $\eta$ and $L_f$ are the
viscous scale and large scale of the system, respectively. It is worth
noticing that such a prediction, which results also from dimensional
analysis \cite{DDM79}, leads to the Kolmogorov spectrum $E(k)\sim
k^{-5/3}$ for $y=4$, i.e.  quite far from the perturbative region
where the RG calculations are under control. Here we want to study
the same problems addressed for the linear case, i.e.  whether
fluctuations are sensitive to the injection mechanism for any $y$ and
if one observes anomalous scaling for $y < 4$, when the forcing is
dominant.  Since, at least with the present knowledge, RG perturbative
methods starting at $y \sim 0$ cannot control anomalous corrections,
only numerical simulations at finite $y$ values, can possibly give
some answers.

In Ref.~\cite{SMP98} a first numerical investigation of $3d$
incompressible Navier-Stokes problem was performed. The authors made a
set of DNS, varying the spectrum slope from $y=3$ to $y=8$ at low
Taylor's Reynolds number $Re_{\lambda}=22$. Without considering the
issue of anomalous scaling in the region $y\,<\,y_c=4$, they mostly
concentrated on the cases with $y \ge 4$, and ended up with the
conclusion that for $y \ge 4$ the properties of the statistics are not
universal, but varies with the forcing spectrum slope. It is however
difficult to consider these as conclusive results, because of the
large error bars affecting the data. Later, in Ref.~\cite{BLT03},
another numerical study of the same problem at a much higher resolution
(up to $Re_{\lambda} =220$) has given some support, even if mostly based on
semi-quantitative results, to the scenario drawn in the previous
section in the context of passive transport.
\begin{figure}[t!]
\begin{center}
\fl \hspace{1truecm} {\includegraphics[draft=false,scale=0.65,clip=true]{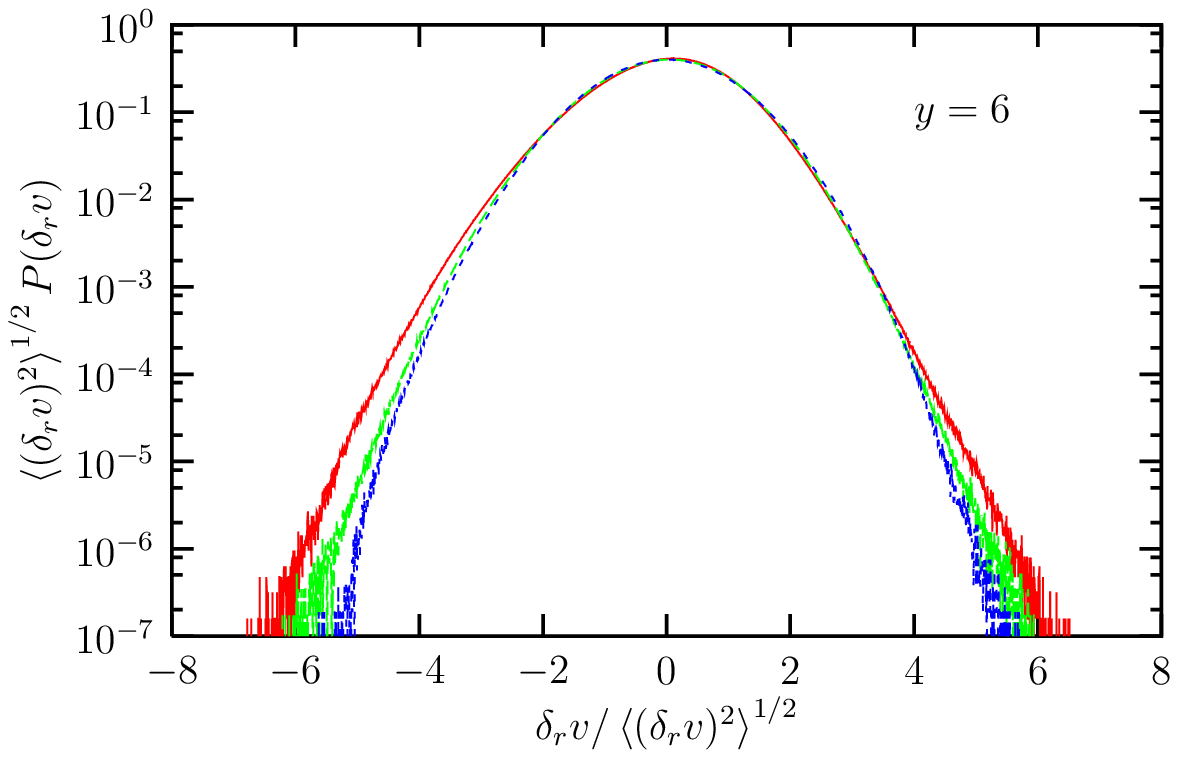}\includegraphics[draft=false, scale=0.65,clip=true]{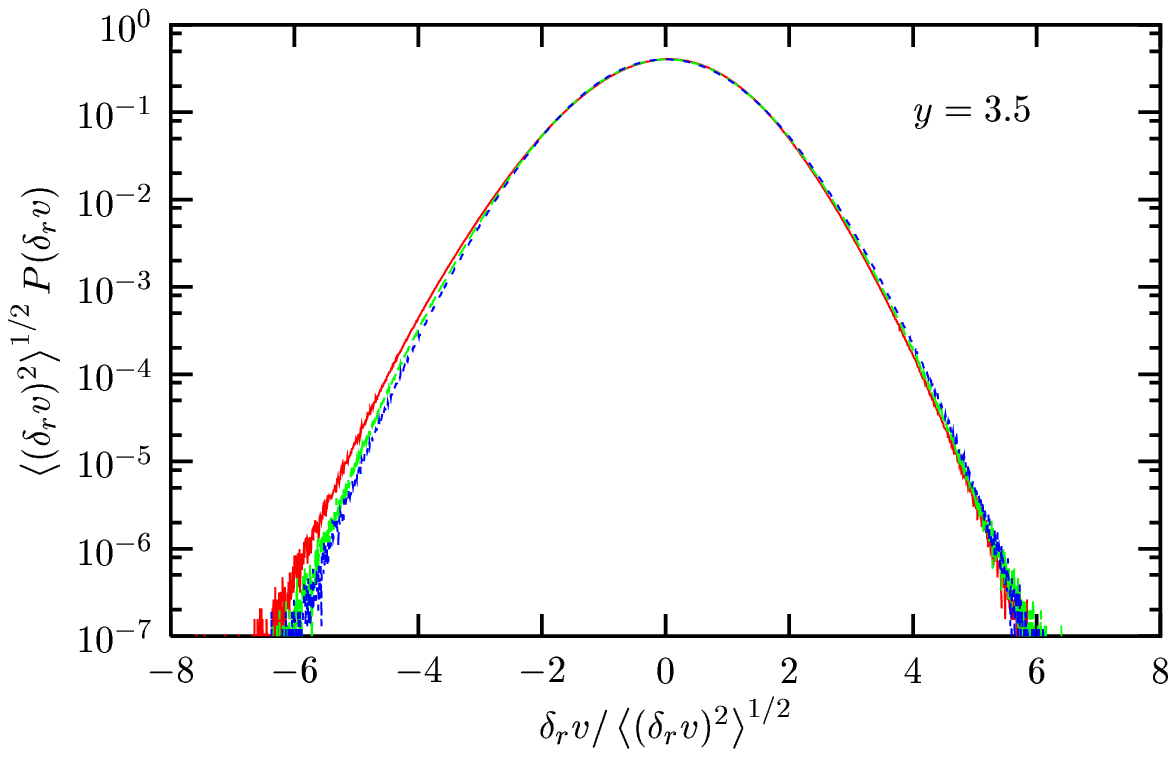}}
\end{center}
\caption{Left: PDF of the velocity increments, for $y=6.0$, for three
separations $r_1=34\eta$ and $r_2=74\eta$ in the inertial range, and
$r_3=114\eta$ in the energy containing range. Distributions are
normalised to have unit variance. Right: the same as in the left
panel, but for the case $y=3.5$. The incompressible Navier-Stokes
equations are solved with a second-order hyper-viscous dissipative
term $\propto \nu \Delta^2$. Temporal integration has been carried
over for about $20-30$ large-eddy turnover times. DNS data refer
to $256^3$ simulations.  The range of the forcing, in Fourier space,
extends down to the maximum resolved wavenumber.  }
\label{fig:6} 
\end{figure} 
In particular, we report here the behaviour of longitudinal velocity
increments PDFs for two among the different runs done in \cite{BLT03},
with $y=3.5$ and $y=6$. For $y=6>y_c$, as shown in Fig.~\ref{fig:6}
left panel, the usual intermittent behaviour for the PDFs is found:
the curves $P(\delta_r v)$ at different separations $r$ do not rescale
one onto the other. On the contrary, for $y=3.5<y_c$, the PDFs at
various scales are almost indistinguishable (see Fig.~\ref{fig:6},
right panel): a signature of the absence of intermittent corrections
at least for the core of the distribution.  It is worth noticing that
this result is far from trivial, because it coincides with the RG
predictions \cite{FNS77,FF78,DDM79,YO86,AAKV03} in the region where RG
calculations are well beyond their range of validity.  These findings,
even if qualitative, fit rather well with those of the passive scalar
case (see Fig.~\ref{fig:4}).

 We may try to push forward this indication in the light of the theory
for linear systems. For the $3d$ Navier-Stokes dynamics, the
stationary equations for multi-point correlators $ {\cal C}_p({{\bm
r}},t)\equiv \langle v_{\alpha_1}({\bm r}_1) v_{\alpha_2}({\bm r}_2)
\ldots v_{\alpha_p}({\bm r}_p)\rangle $ can be sketched as,
\begin{equation}
\label{eq:closure}
\Gamma_{p+1} \,{\cal C}_{p+1} + \nu \,D_{p}\,{\cal C}_{p} +
{\cal F}_{2} {\cal C}_{p-2}=0,
\end{equation} 
where $\Gamma_{p+1}$ is the integro-differential linear operator
coming from the inertial and pressure terms, $D_{p}$ is the
differential operator describing dissipative effects and ${\cal
F}_{2}$ is the two-point forcing correlator. Unfortunately, since the
hierarchy (\ref{eq:closure}) is unclosed, no
straight-forward analytical approache can be done.  
Even though nothing can be rigorously said, one may still be tempted
to assume that anomalous scaling is brought by the inertial
operator. However, the presence of finite size effects, due to the
limited inertial range, in DNS data of $3d$ Navier-Stokes simulations
\cite{BLT03} provide only a semi-quantitative support to this
scenario. As we shall see in the next section, more quantitative
statements can be made in the context of shell models for
turbulence, for which resolution constraints are much less severe.
\subsection{Shell models for turbulence}
\label{sec:shell}
Shell models of turbulence are dynamical systems mimicking
Navier-Stokes nonlinear evolution \cite{BJPV98}. Their main advantage
relies on the possibility of performing high Reynolds number
simulations to measure statistical properties, say intermittent
corrections, with high accuracy. The model we use, proposed in
\cite{LPPPV96}, is an improved version of the GOY model
\cite{YO87,JPV91} (see also \cite{B03} for a recent review). The
evolution equation is\,:
\begin{equation}
\label{eq:shell} 
\fl \frac{d}{dt}u_{n} =i(
k_{n}u^{*}_{n-2}u_{n-1}+b\,k_{n-1}u^{*}_{n-1}u_{n+1}+(1+b)
k_{n-2}u_{n+1}u_{n+2}) -\nu k^{2}_{n}u_{n}+f_{n}\,.
\end{equation}
The velocity field fluctuation at the wavenumber $k_{n}$, with
$k_{n}=2^{n}k_{0}$, is expressed in terms of the complex variable
$u_{n}$; $b$ is a free parameter and $\nu$ indicates the viscosity.
The number of shells varies as $n=\{0,\dots,N_{max}\}$.  Remarkably,
for large scale forcing, the structure functions show
anomalous scaling:
\begin{equation}
\label{def:shellanomalo}
S_{p}(k_{n})\equiv \langle |u_{n}|^{p}\rangle \sim k^{-\zeta_p}_{n}
\end{equation} 
deviating from the dimensional prediction $\zeta_{p}=p/3$. Moreover,
anomalous scaling exponents are found to be universal with respect
to the forcing, provided it is large scale (see \cite{B03}). 

We aim at investigating the properties of the model stirred by a
white-in-time, Gaussian field, with zero mean and spectrum
\begin{equation}
\label{def:shellforce}
\langle |f_{n}|^2 \rangle = f_0 k^{\beta}_{n},
\end{equation}   
being $f_0$ the forcing intensity.  A similar injection mechanism was
proposed in \cite{MOV02} where authors tried to compare the power-law
forced shell models with fractal-grid induced turbulence \cite{vas2}.
In order to understand the effect of forcing as (\ref{def:shellforce})
let us start with the energy flux, which obeys an exact
equation. Denoting with $\dot{E}_{N} ={d \over {dt}}
\sum_{n=0}^{N}\langle {|u_{n}|^{2}}/{2} \rangle$ the time derivative
of the energy content up to the scale $N$, the energy balance relation
can be written as :
\begin{equation} 
\label{eq:shellenergy}
\dot{E}_N+\nu
\sum_{n=0}^{N}k^{2}_{n}\langle |u^{2}_{n}|\rangle = 
-k_{N}\Pi_{N}+\sum_{n=0}^N \langle |f_{n}|^2 \rangle\,.
\end{equation}
In the previous expression, $\Pi_{N}= \langle {\cal T}_N\rangle$,
where ${\cal T}_N$ is the energy flux through the shell $N$ defined as
\begin{equation}
\label{def:shellflux}  
{\cal T}_N =\Im \left\{u^{*}_{N}u^{*}_{N+1}u_{N+2} + \frac{(1+b)}{2}
u^{*}_{N-1}u^{*}_{N} u_{N+1}\right\}\,.
\end{equation}
For the forcing (\ref{def:shellforce}) and for $\nu\to 0$, we get an
exact prediction in the stationary state:
\begin{equation}
\label{eq:shellflusso}
\Pi_{n}=A_1 k^{-1}_{n}+B_1 k^{-1+\beta}_{n}\,,
\end{equation}  
where the constants $A_1,B_1$, bound by the equation of motion to have
opposite signs, depend on the forcing details ($f_0$, $\beta$) and on
the integral wavenumber $k_{0}$. At this level, where everything is
exact, we see that when $\beta$ approaches the critical value
($\beta=0$), a large number of shells is needed to distinguish leading
terms from subleading ones in the third order moment. In our runs, the
choice of a large value for $N_{max}(=40)$, allow us to have a good
control of such effects.

\begin{figure}[t!] 
\begin{center}
{\includegraphics[draft=false,
scale=0.47,clip=true]{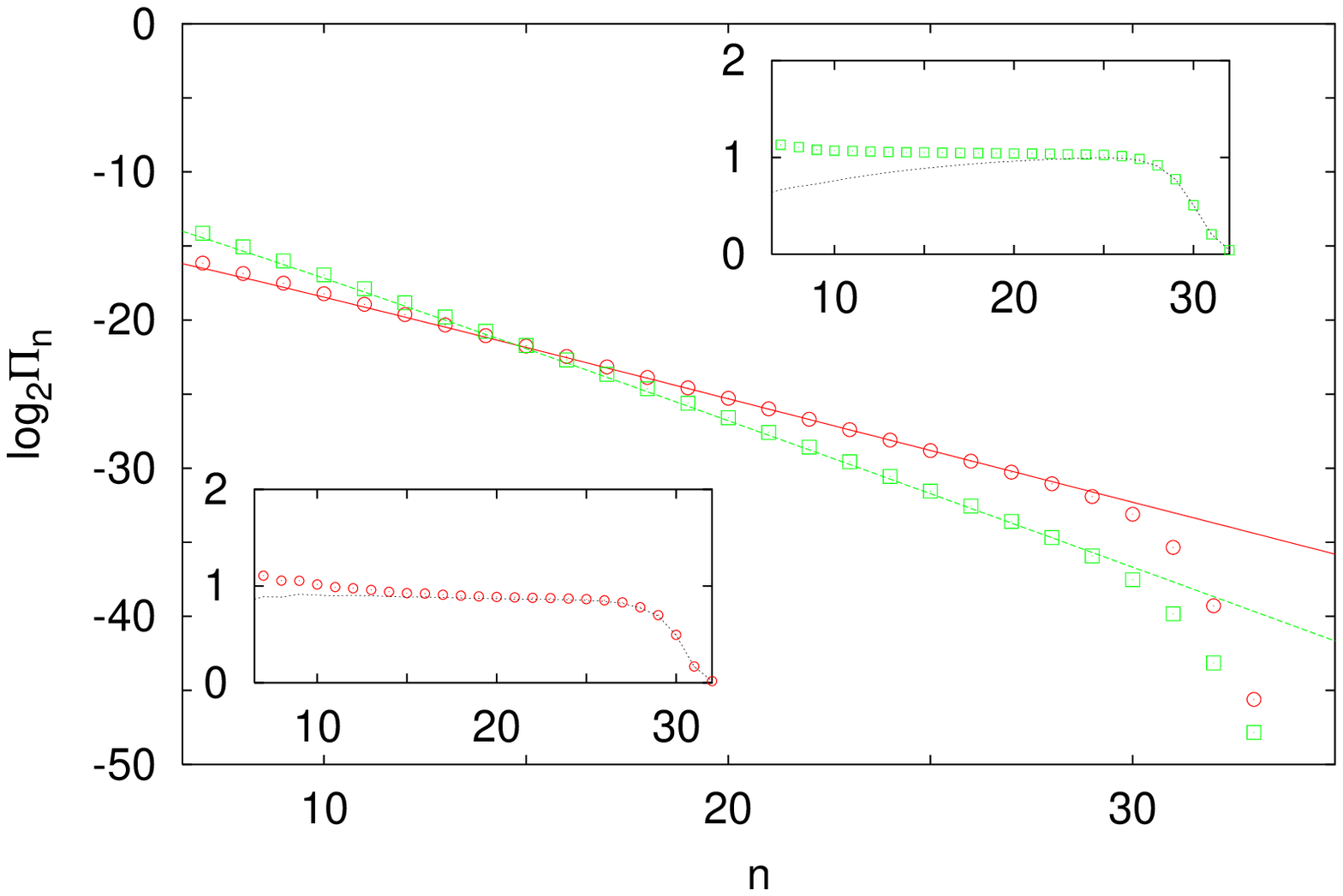} 
\includegraphics[draft=false, scale=0.47,clip=true]{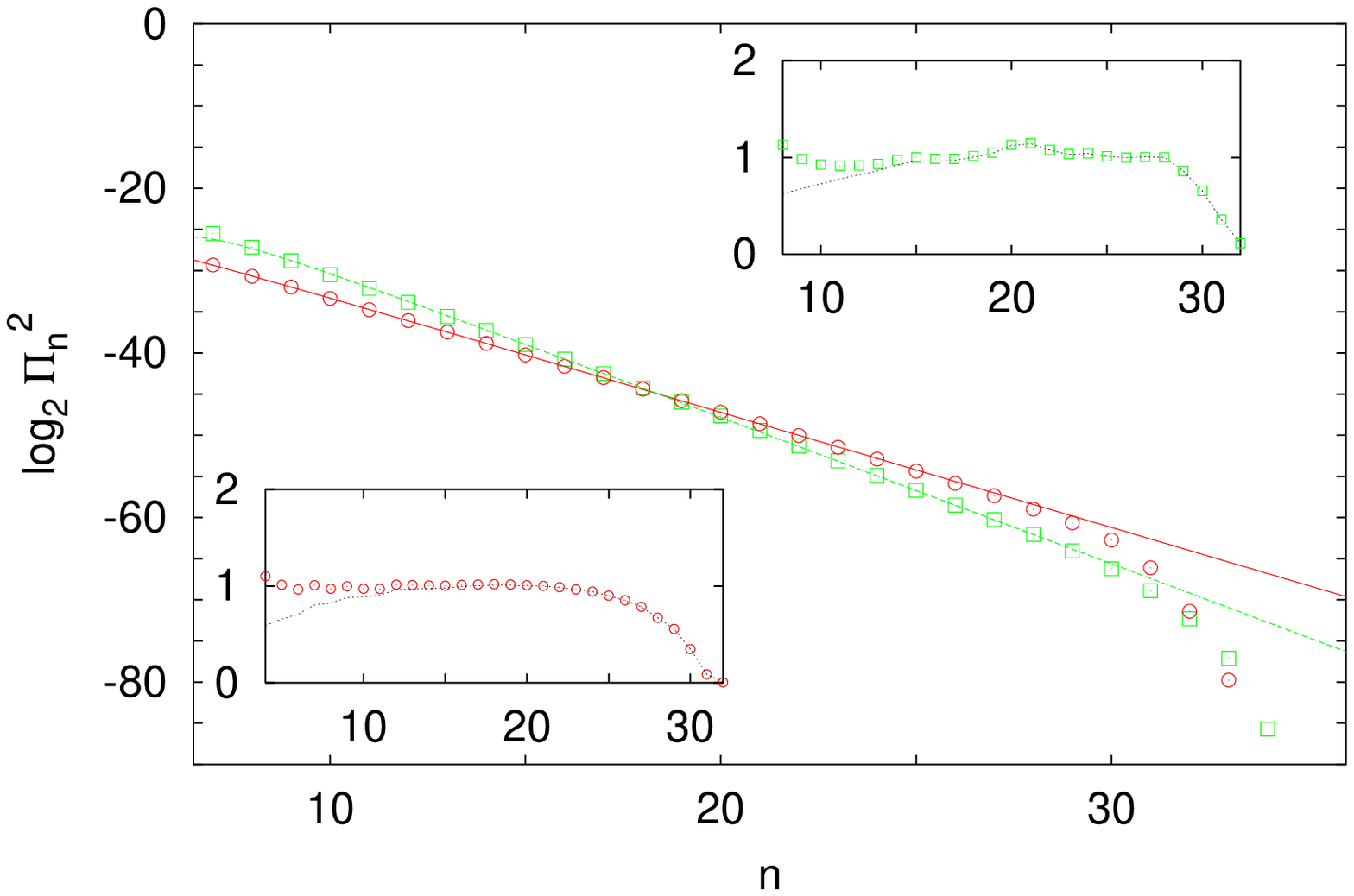}}
\end{center}
\caption{Left: Shell fluxes measured in the presence of a power-law
forcing. In our runs, we adopted the following choice of parameters:
$b=-0.4$, $\nu=10^{-12}$, $N_{max}=40$. For the random forcing, acting
on all shells of the inertial range, we chose the following values of
$\beta$: $\beta=0.3$ (green squares), $\beta=-0.15$ (red
circles). Fits are made with a superposition of two power-law
contributions, in agreement with (\ref{eq:shellflusso}). Insets:
compensation with the leading power law (dotted) or with two power
laws for $\beta=0.3$ (bottom inset) and $\beta=-0.15$ (top inset).
Right: same as left panel but for the second order flux moment, to be
compared with (\ref{eq:sub}). Also here the two power laws terms have
opposite signs.}
\label{fig:7} 
\end{figure}

Similarly to the NS case (\ref{eq:closure}), we can write the 
 unclosed hierarchy for the multi-point correlators of shell variables:
\begin{equation}
\label{shell:eqs}
\frac{d}{dt}C_{p}(\underline{n})=
M_{p+1}(\underline{n},\underline{n}') \,C_{p+1}(\underline{n}')+
F_{2}(\underline{n},\underline{n}') \,C_{p-2}(\underline{n}')\,,
\end{equation}
where $C_{p}(\underline{n})=\langle u_{n_{1}}u_{n_{2}}...u_{n_{p}}
\rangle $ is the generic correlation function of order $p$,
$M_{p+1}(\underline{n},\underline{n}')$ is the linear operator coming
from the inertial terms, and $F_{2}(\underline{n},\underline{n}')$ is
the two-point forcing operator. The main difference with the case of
passive scalar transport, as pointed out in the previous subsection,
is that now the system (\ref{shell:eqs}) is not closed, i.e. we have
more unknowns than equations. Without any ambition of a rigorous
approach, we may imagine that the general solution is characterised by
two terms. The first, $C^{{\cal I}}_{p}$, given by the dimensional matching
between the inertial operator and the forcing operator,
\begin{equation}
\label{eq:ino}
M_{p+1}(\underline{n},\underline{n}')\, C^{\cal
I}_{p+1}(\underline{n}')=
F_{2}(\underline{n},\underline{n}')\,C^{\cal I}_{p-2}(\underline{n}').
\end{equation}
The second, ${\cal Z}_{p}$, associated to a ``zero mode'' of the
inertial operator, namely a solution of the homogeneous equation,
$M_{p+1}(\underline{n},\underline{n}') {\cal
Z}_{p+1}(\underline{n}')=0\,$. Here, as for the linear problem, the
non homogeneous terms are the only contributions depending on the
forcing, while the homogeneous ones, dimensionally unconstrained, may show
anomalous scaling. In particular, for the structure functions
$S_{p}(k_n)$, we have
\begin{equation}
\label{solution} 
S_{p}(k_n) \sim k^{-\zeta_{p}}_{n} + k^{-\zeta^{dim}_p(\beta)}_{n}\,,
\end{equation} 
where $ \zeta^{dim}_p(\beta) = p/3(1-\beta)$.  This implies the
following behaviour for the second order moment of the flux
(equivalent to the sixth-order structure function):
\begin{equation}
\label{eq:sub}
\Pi^2_{n}= \langle {\cal T}_n^2 \rangle = A_2 k^{-1.8}_{n}+ B_2
k^{2/3(\beta-1)}_{n},
\end{equation}
where the power $1.8$ coincides with the anomalous exponent measured
for sixth-order structure function with a large scale forcing, and the
value $2/3(\beta-1)$ comes from the dimensional prediction (\ref{eq:ino}).

In Fig.~\ref{fig:7} the first and second moments of the flux variables
are shown for both the subleading and dominant forcing with
$\beta=-0.15$ and $\beta=0.3$, respectively. It is easy to see that to
get a good agreement of the numerical results with the prediction
(\ref{eq:sub}), both the leading and the subleading power laws have to
be taken into account.
For both cases with $\beta=0.3$ and $\beta=-0.15$, at least for the
order of the statistics we could reach, the anomalous scaling
exponents have values in agreement with those found for a large scale
forcing, supporting the universality scenario.

\section{Conclusions}
\label{sec:conclude}
We have discussed the problem of small scale fluctuations in turbulent
systems stirred at all scales by a power-law forcing. The main
question in our study concerns anomalous scaling and universality of
scaling exponents.

For linear systems -- i.e. passive transport--, we find clear evidence for
the universality of anomalous fluctuations and our results fit well in
the zero mode scenario. In the regime in which forcing is subleading
anomalous scaling is recovered in quantitative agreement with the case
of large scale forcing. In the forcing dominated regime, the
dimensional scaling imposed by the injection mechanism overwhelms the
anomalous fluctuations of low order moments. Nevertheless, anomalous
scaling shows up again for high enough moments, independently
of the forcing spectrum slope, as confirmed by the existence of a
unique saturation exponent in all regimes.

The results obtained for nonlinear turbulent systems point in the same
direction. Indeed both the semi-quantitative results obtained in the
context of $3d$ Navier-Stokes turbulence and those more quantitative
issuing from the investigation of the shell models are compatible with
the linear systems scenario for universality.  However, in the
presence of power-law forcing acting in the turbulent energy cascade
range, as shown both in the linear and nonlinear cases, scaling
properties are strongly spoiled by the beating of leading and
subleading terms. This effect is particularly strong due to
cancellations induced by different signed pre-factors in the power-law
terms.

Strong cancellation effects, which apparently are a mere technical
question, may lead to misinterpretation in analysing data.
We wonder, for example, if the observed multi-fractal
behaviour in the one-dimensional Burgers equation stirred by a power-law
forcing might be a spurious effect \cite{HJ97}.

\ack
We are grateful to J.~Bec, A.~Celani and U.~Frisch for useful
discussions.  We acknowledge J.~Davoudi who participated to the early
stage of the study for passive scalar problem.  This research was
partially supported in part by the EU under the grant No. HPRN-CT
2000-00162 ``Non Ideal Turbulence'' and by the INFM (Iniziativa di
Calcolo Parallelo). A.~L. acknowledges A.~Pouquet for useful
discussions when visiting NCAR (Boulder, Colorado), and CNR for the
support received under the grant ``Short-term mobility of researchers
2003''.


\section*{References}
\begin{thebibliography}{99} 
\bibitem{frisch} Frisch U 1995 
                 {\it Turbulence: The Legacy of A.~N.~Kolmogorov} 
                 (Cambridge: Cambridge University Press)

\bibitem{NWLMF97} Noullez A, Wallace G, Lempert W, Miles R B and Frisch U 1997 
                  {\it J. Fluid Mech.} {\bf 339}, 287

\bibitem{LPVCAB01} La Porta A, Voth G A, Crawford A M, Alexander J, Bodenschatz E 2001 
                   {\it Nature} {\bf 409}, 1017

\bibitem{MMMP01} Mordant N, Michel O, Metz P and Pinton J F 2001
                 {\it Phys. Rev. Lett.} {\bf 87}, 21

\bibitem{W00} Warhaft Z 2000 
              {\it Annu. Rev. Fluid Mech.} {\bf 32}, 203

\bibitem{GFN02} Gotoh T, Fukayama D and Nakano T 2002
                {\it  Phys. Fluids} {\bf 14}, 1065

\bibitem{BCV00} Boffetta G, Celani A and Vergassola M 2000
                {\it Phys. Rev. E} {\bf 61}, R29 

\bibitem{SA97} Sreenivasan K R and Antonia R A, 1997
               {\it Annu. Rev. Fluid Mech.} {\bf 29}, 435

\bibitem{ABBBC96}Arneodo A, Baudet C, Belin F, Benzi R, Castaing B, et al. 1996 
                 {\it Europhys. Lett.} {\bf 34}, 411


\bibitem{ADKLPS98}Arad I, Dhruva B, Kurien S,  L'vov V S, Procaccia I and Sreenivasan K R 1998  
                  {\it Phys. Rev. Lett.} {\bf 81}, 5330


\bibitem{KS00} Kurien S and Sreenivasan K R 2000
               {\it Phys. Rev. E} {\bf 62}, 2206

\bibitem{SW02} Shen X and Warhaft Z 2002
               {\it Phys. Fluids} {\bf 14}, 2432

\bibitem{BT00} Biferale L and Toschi F 2001
               {\it  Phys. Rev. Lett.} {\bf 86}, 4831

\bibitem{BCTT03}Biferale L, Calzavarini E, Toschi F and Tripiccione R 2003 
                {\it Europhys. Lett.} {\bf 64}  461

\bibitem{K68} Kraichnan R H 1968 
              {\it Phys. Fluids} {\bf 11}, 945

\bibitem{FGV01} Falkovich G, Gaw\c{e}dzki K and Vergassola M 2001
                {\it  Rev. Mod. Phys.} {\bf 73}, 913

\bibitem{CV01}Celani A and Vergassola M 2001 
              {\it Phys. Rev. Lett.} {\bf 86}, 424

\bibitem{CLMV01} Celani A, Lanotte A, Mazzino A and Vergassola M 2001
                 {\it Phys. Fluids} {\bf 13}, 1768

\bibitem{ABCPV01}Arad I, Biferale L, Celani A, Procaccia I and Vergassola M 2001
                 {\it Phys. Rev. Lett.} {\bf 87}, 164502

\bibitem{FNS77} Forster D, Nelson D R and Stephen M J 1997
                {\it Phys. Rev. A} {\bf 16} 732 

\bibitem{FF78} Frisch U and Fournier J D 1978
               {\it  Phys. Rev. A} {\bf 17}, 747 

\bibitem{DDM79}De Dominicis C and Martin P C 1979
               {\it  Phys. Rev. A} {\bf 19} 419

\bibitem{YO86}Yakhot V and Orszag S A 1986
              {\it  Phys. Rev. Lett.} {\bf 57}, 1722 

\bibitem{AAKV03}Adzhemyan L Ts, Antonov N V, Kompaniets M V and Vasil'ev A N 2003 
                {\it Intern. Journ. Modern Physics B} {\bf 17}, 2137 

\bibitem{EY94}Eyink G 1994 
              {\it Phys. Fluids} {\bf 6}, 3063

\bibitem{SMP98}Sain A, Manu and Pandit R 1998
               {\it Phys. Rev. Lett.} {\bf 81}, 4377

\bibitem{BLT03} Biferale L, Lanotte A and Toschi F 2004
                {\it Phys. Rev. Lett.} to appear; 
                e-Print arXive nlin.CD/0310022

\bibitem{GK95} Gaw\c{e}dzki K and Kupiainen A 1995 
               {\it  Phys. Rev. Lett.} {\bf 75}, 3834

\bibitem{CFKL95} Chertkov M, Falkovich G, Kolokolov I and Lebedev V 1995
                 {\it Phys. Rev. E} {\bf 2}, 4924

\bibitem{BL98} Balkovsky E and Lebedev V 1998
               {\it Phys. Rev. E}  {\bf 58}, 5776

\bibitem{MWAT01}Moisy F, Willaime H, Andersen J S and Tabeling P 2001
                {\it  Phys. Rev. Lett.} {\bf 86}, 4827 

\bibitem{BGK98} Bernard D, Gaw\c{e}dzki K and Kupiainen A 1998
                {\it J. Stat. Phys.} {\bf 90}, 519 

\bibitem{PT98} Paret J and Tabeling P 1998
               {\it  Phys. Fluids} {\bf 10}, 3126 

\bibitem{T02}Tabeling P 2002 
             {\it Phys. Rep.} {\bf 362}, 1 

\bibitem{MY75} Monin A and Yaglom A 1975 
               {\it Statistical FluidMechanics}, Vol.2 
               (Cambridge: MIT Press)

\bibitem{BJPV98} Bohr T, Jensen M H, Paladin G and Vulpiani A 1998
                 {\it Dynamical system approach to turbulence} 
                 (Cambridge: Cambridge University Press)

\bibitem{LPPPV96}L'vov V S, Podivilov E, Pomyalov A, Procaccia I and Vandembroucq D 1996 
                 {\it Phys. Rev. E} {\bf 58}, 1811 

\bibitem{YO87}Yamada M and Ohkitani K 1987 
              {\it J. Phys. Soc. Jpn.} {\bf 56}, 4210 

\bibitem{JPV91}Jensen M H, Paladin G and Vulpiani A 1991
               {\it Phys. Rev. A} {\bf 43}, 798 

\bibitem{B03}Biferale L 2003 
             {\it Ann. Rev. Fluid Mech.} {\bf 35}, 441

\bibitem{MOV02}Mazzi B, Okkels F and Vassilicos J C 2002 
               {\it Eur. Phys. J. B} {\bf 28}  243

\bibitem{HJ97} Hayot F and Jayaprakash C 1997
               {\it Phys. Rev. E} {\bf 56}, 4259


\bibitem{vas2} Staicu A, Mazzi B,  Vassilicos J C and van de Water W 2003
               {\it Phys. Rev. E} {\bf 67}, 066306
\end{thebibliography}
\end{document}